# Diffusion MRI measurements in challenging head and brain regions via cross-term spatiotemporally encoding

Eddy Solomon[a], Gilad Liberman[a], Zhiyong Zhang and Lucio Frydman*

*Department of Chemical Physics, Weizmann Institute of Science, Rehovot 76100, Israel*

**Word count:** 4,309 in main text (including Methods and Captions), 1,138 in supporting information and extended data

**Figures:** 6 in main text, 2 in supporting information and extended data

**Tables:** 0

**Equations:** 2 in main text, 4 in supporting information and extended data

**References:** 48 in main text, 11 in supporting information and extended data

[a] Contributed equally to this study

*Prof. Lucio Frydman; +972-8-9344903; lucio.frydman@weizmann.ac.il

## Summary

Cross-term spatiotemporal encoding (xSPEN) is a recently introduced imaging approach delivering single-scan 2D NMR images with unprecedented resilience to field inhomogeneities. The method relies on performing a pre-acquisition encoding and a subsequent image read out while using the disturbing frequency inhomogeneities as part of the image formation processes, rather than as artifacts to be overwhelmed by the application of external gradients. This study introduces the use of this new single-shot MRI technique as a diffusion-monitoring tool, for accessing regions that have hitherto been unapproachable by diffusion-weighted imaging (DWI) methods. In order to achieve this, xSPEN MRI's intrinsic diffusion weighting effects are formulated using a customized, spatially-localized *b*-matrix analysis; with this, we devise a novel diffusion-weighting scheme that both exploits and overcomes xSPEN's strong intrinsic weighting effects. The ability to provide reliable and robust diffusion maps in challenging head and brain regions, including the eyes and the optic nerves, is thus demonstrated in humans at 3T; new avenues for imaging other body regions are also briefly discussed

# Introduction

Pulsed field gradients enable one to measure water's diffusivity under *in vivo* conditions (1-4). By measuring the extent and directionality of this microscopic property, diffusion-weighted and diffusion-tensor imaging (DWI, DTI) provide unique opportunities for extracting structural insights about tissues in general, and about human organ morphology in particular (2,5,6). Single-scan "ultrafast" magnetic resonance imaging (MRI) methods play a primary role in such *in vivo* studies, as in multi-scan sequences motions will generally interfere with gradients employed in the diffusion- and the position-encoding stages, corrupting the quantitativeness of the experiments (7,8). Spin-echo echo-planar imaging (SE-EPI) techniques capable to deliver quality 2D imaging information in a single scan are therefore widely used in preclinical and clinical diffusion studies (9). Despite SE-EPI's critical role in these studies (10-12), this sequence is also prone to artifacts that limit its use to study diffusion of relatively large, homogeneous organs. Particularly deleterious are the effects of external field ($\Delta B_o$) and chemical shift heterogeneities, which can introduce severe geometrical distortions and corrupt the diffusivity arising from these maps (13,14). A number of alternatives based on radial and multi-spin-echo sampling strategies, have been proposed for tackling these limitations (15-19). Spatiotemporal encoding (SPEN) principles provide yet another alternative to achieve an enhanced immunity to these distortions under similar acquisition, sensitivity and resolution parameters conditions (20-23). Spin-packets in SPEN echo throughout the course of the acquisition rather than at a single time as in conventional spin echoes. It has been shown that this makes out of SPEN a robust tool to map diffusion in preclinical (24,25) and clinical settings (26,27) as well as in functional MRI (28,29), under conditions leading to severe image distortions from SE-EPI. Recently we have introduced an alternative modality dubbed cross-term SPEN (xSPEN, 30), which is not only endowed with these robust echoing capabilities, but also free from $\Delta B_o$- or chemical-shift-driven misplacements. xSPEN provides this unprecedented resilience to field heterogeneities by recruiting –rather than overcoming– field inhomogeneities into the image encoding and decoding processes. To do so xSPEN relies on a continuous background gradient, which adds as a source of line broadening to whatever field or shift distortions exist. In combination with frequency-swept pulses imparting a saddle-shaped phase profile onto the spin-packets (30,31), a stationary-phase focal point is created that over the course of an acquisition performed under the constant action of the same inhomogeneous broadenings displaces over the object and thereby rasterizes the sought profile –delivering it free from distortions.

This works exploits these unique xSPEN capabilities, to map diffusion in areas that have hitherto proved inaccessible to single-shot studies. This mapping is complicated by the constant gradient employed by xSPEN MRI, which while capable to provide $T_2^*$-free images for a large range of $\Delta B_o$ inhomogeneities, also imparts a heavy diffusion weighting. Such weightings challenge the monitoring of motions along orthogonal axes, making it hard to obtain full tensorial –or even isotropically-weighted– diffusion information. In order to perform such full tensorial mapping this study relies on a local *b*-matrix analysis, and uses it to devise xSPEN-based pulsed-gradient spin-echo (PGSE) schemes that can sample diffusivity over a sufficiently large range of *b*-space orientations. The usefulness of these new tools is demonstrated in a series of preclinical and clinical imaging tests providing diffusivity information in human head and brain regions that are often unreachable by single-shot methods.

**Results**

*xSPEN MRI*. To better understand the challenges that xSPEN imaging poses to diffusion measurements, it is convenient to briefly review the features that distinguish this methodology from its SPEN predecessors. SPEN relies on a progressive excitation/inversion and refocusing of the spins, achieved by applying a frequency-swept radio frequency (RF) pulse lasting for a time $T_e$ and acting whilst in the presence of an encoding gradient $G_e$ (32,33). Whether this RF pulse is used for an excitation or inversion, the result is a spatially parabolic phase profile. Assuming for concreteness swept $180^0$ inversion pulses applied while in the presence of a *y*-axis gradient, this phase can be written, within an unimportant constant, as

$$\varphi_e(y) = -\frac{(\gamma G_e)^2}{R} y^2 \qquad [1]$$

where $R$ is a sweep rate defined by $\gamma G_y FOV/T_e$, and *FOV* defines the targeted field-of-view along the y-axis being encoded. The quadratic coefficient of the parabolic phase in Eq. [1] defines the spatial extent of the spins emitting at any given moment, as signal emission will be dominated by spins positioned at the apex of this phase parabola. To probe the full *FOV* this stationary point is displaced, by applying an additional acquisition gradient $G_a$ over an acquisition time $T_a$. Extending this 1D rasterization into a single-shot 2D MRI experiment requires encoding a second (e.g., *x*-) axis, something that is usually achieved by a conventional oscillating readout (RO) gradient. Fourier transform (FT) along the RO axis followed by a magnitude or super-resolved calculation, then delivers the final 2D image (21,34,35).

The xSPEN pulse sequence takes Eq. [1] one step further by melding into it, the kind of encoding used in in so-called "ultrafast" single-shot 2D NMR spectroscopy (36,37). In this experiment the initial spin excitation is combined with two frequency-swept inversion pulses, acting in unison with a pair of bipolar gradients. This replaces the parabolic phase introduced in Eq. [1] by a bilinear phase encoding, proportional to both the chemical shifts $\Omega_i$ of the targeted sites as well as on the spins' positions along the axis of the bipolar gradient: $\varphi_e = C\Omega_i y + \varphi_0$, with $C=4T_e/FOV$ a spatiotemporal constant under control and $\phi_0$ a position-independent phase. Having imposed such encoding, the application of $G_a$ during acquisition leads to the generation of site-specific gradient echoes, enabling the acquisition of arbitrary nD NMR correlations in a single scan. In xSPEN, this spectroscopy-oriented approach is converted onto an imaging one by replacing the chemical shifts $\Omega_i$s by a spatially-dependent frequency. This inhomogeneous frequency broadening can be imparted, for instance, by activating a constant $G_z$ along the slice-select z-axis (Figure 1a). Considering an unknown term $\delta\omega(r)$ coming from field heterogeneities and chemical shifts that adds to this z gradient, leads to an encoding phase profile

$$\varphi_e(y,z) = -Cy \cdot [\gamma G_z z + \delta\omega(y,z)] \qquad [2]$$

$C = \dfrac{4T_e \gamma G_y}{2\pi BW}$ is now defined as a function of the bandwidth *BW* characterizing the swept RF that is applied while in the presence of the two encoding gradients: $BW = (\gamma G_y FOV_y + \gamma G_z L_z)/2\pi$, with $L_z$ a nominal slice thickness. The hyperbolic encoding phase *yz* dominating Eq. [2] has the unique feature that it allows the sum of the "encoding" frequencies $\gamma G_z z + \delta\omega(y,z)$ to eventually decode a distortion-less *y* image. In other words, $G_z$ and $\delta\omega$ become both the encoding mechanism of the experiment, as well as the decoding tool revealing the undistorted positions of the spins along the *y*-axis –regardless of the size of the field distortion (30). Figure 1c exemplifies this by presenting results collected on a phantom incorporating a titanium screw and a Lego® piece; clearly xSPEN provides more faithful representations of these objects than any of the remaining single-shot counterparts.

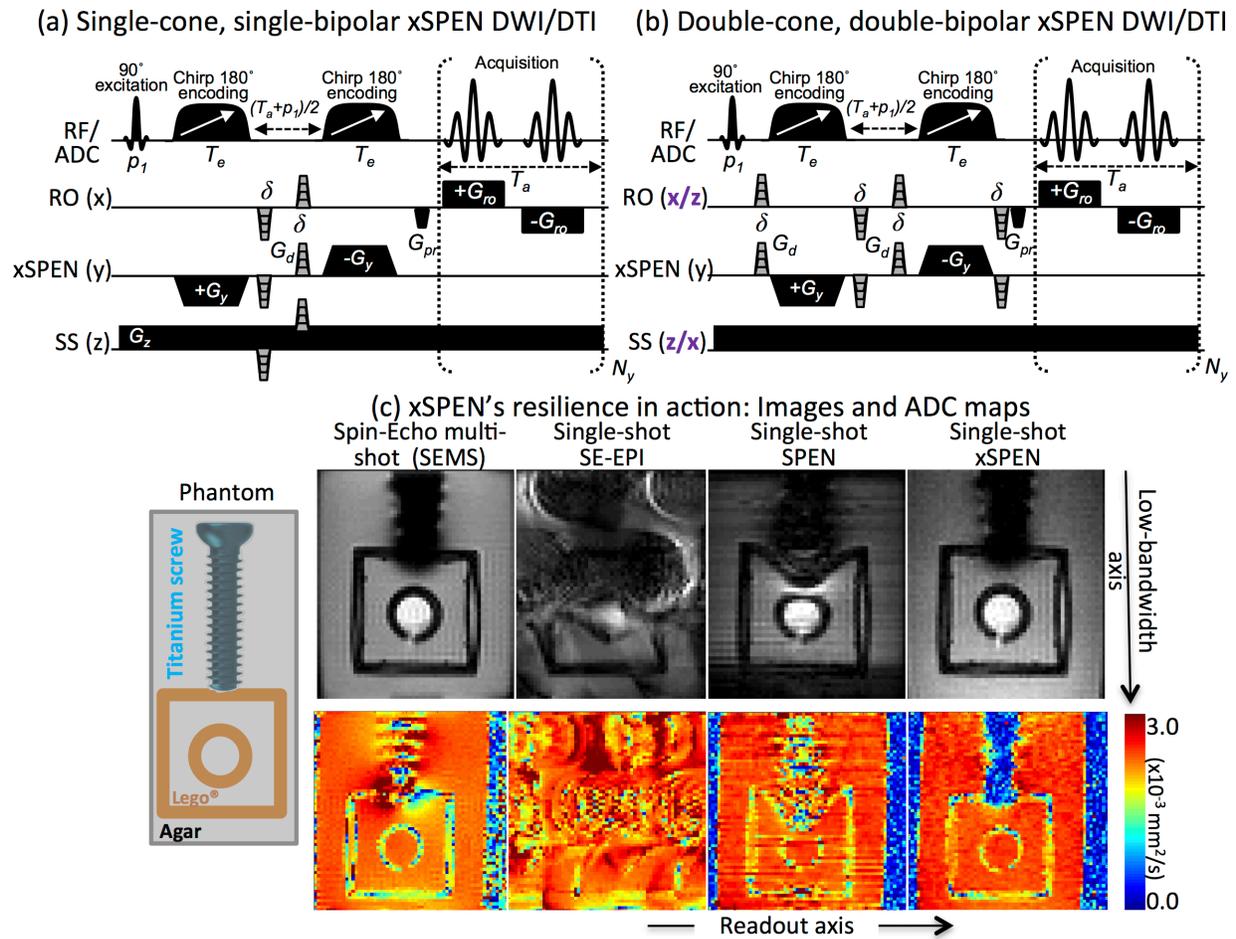

**Figure 1:** (a,b) xSPEN diffusion pulse sequences assessed in this study, involving in all cases a slice-selective 90⁰ excitation pulse ($p_1$) followed by two 180⁰ chirped pulses. In (a), a single diffusion-weighting PGSE block is placed in the pre-encoding $(T_a+p_1)/2$ delay required for targeting the desired FOV under full-refocusing. In (b) two PGSE diffusion blocks are placed on both sides of the 180⁰ chirped pulses to enable a larger diffusion weighting, and both RO and SS axes are alternated among $x$ and $z$ orientations in order to overcome the otherwise dominating $b_{zz}$-weighting derived from the $G_z$ gradient. The RF/ADC line displays the pulses and signal acquisition; RO, SPEN and SS display orthogonal gradient directions; $G_d$ are diffusion-weighting gradients of duration d and stepped amplitudes (in grey); $G_{pr}$, purge gradients; $G_{ro}$, readout acquisition gradients; $T_a$, acquisition time. (c) MRI results obtained on a preclinical 7T scanner for an agar phantom containing a titanium screw and a Lego® piece arranged as indicated on the left, imaged by a reference multi-shot spin echo and by several (SE-EPI, SPEN, xSPEN) single shot methods. In the latter case, the sequence in panel (a) was used with $G_d$=0. Notice that while metal-induced field distortions are evident in all four methods, the xSPEN results most faithful reproduce the original phantom distribution and provide the most reliable ADC map.

*Alternative diffusion-monitoring strategies.* The application of a continuous $G_z$ over the course of both the encoding and the decoding processes, imparts a heavy diffusion weighting into xSPEN MRI. This weighting is further complicated by xSPEN's progressive spatial inversion/observation of the spins' throughout the encoding and acquisition processes, which make these diffusion losses dependent on the $y$ position being decoded. Understanding these effects requires a framework capable of computing the diffusion-derived signal attenuation produced by gradients such as $G_z/\delta\omega$ and $G_y$, as well as accounting for the frequency (i.e., spatially) progressive nature of the RF encoding and readout processes (38, 25,26). Supporting Information A describes the formalism that developed to estimate these effects, and to calculate diffusivity maps in single-shot xSPEN MRI. Based on those analytical and numerical derivations, Figure 2 describes in further detail the challenges that the usual, variable-$G_d$ PGSE scheme, faces in the retrieval of xSPEN diffusivity data. Indeed, whereas in EPI the weak effects of the imaging gradients allow one to explore the full range of necessary *b*-values and orientations by suitably varying the strengths and directions of the diffusion gradients (Fig. 2a), the continuous $G_z$ gradient employed in xSPEN strongly biases the range of *b*-elements and orientations that can be sampled. For instance, when considering bipolar diffusion modules whose directions are uniformly spread over the *x-y* plane, the actual experiments end up subtending a cone in the ($b_{xx}$, $b_{yy}$, $b_{zz}$) sub-space –even though solely the first two of these should have differed from zero (Fig. 2b, gray). Details of this behavior are further illustrated in Extended Data Figure S1, which describes how these cones –and other diffusion-related aspects– vary for different positions along the imaged axis over the course of an xSPEN acquisition. These strong *b*-modulations may not impede diffusivity measurements based on xSPEN PGSE schemes if dealing with preclinical scanners, where strong gradients capable of spreading *b*-values over a sufficiently large value are available. Figure 3 illustrates this with axial slices recorded on an *ex-vivo* rat head using xSPEN and SE-EPI pulse sequences using the standard PGSE scheme in Fig. 1a. Both SE-EPI and xSPEN methods provide similar ADC maps for the brains, even if distortions for other regions in the head are noticeable in the EPI (yellow arrows, Fig. 3b). Still, this flexibility will not be available for the ca. ten-fold weaker pulsed gradients available in conventional clinical scanners. Not even the sequence shown in Fig. 1b, incorporating a second bipolar diffusion-weighting block (39) placed on the far ends of both inversion pulses and furnishing higher *b*-values for a given maximal $G_d$ strength, will provide a sufficient range of accessible *b*-values under clinical scanning conditions.

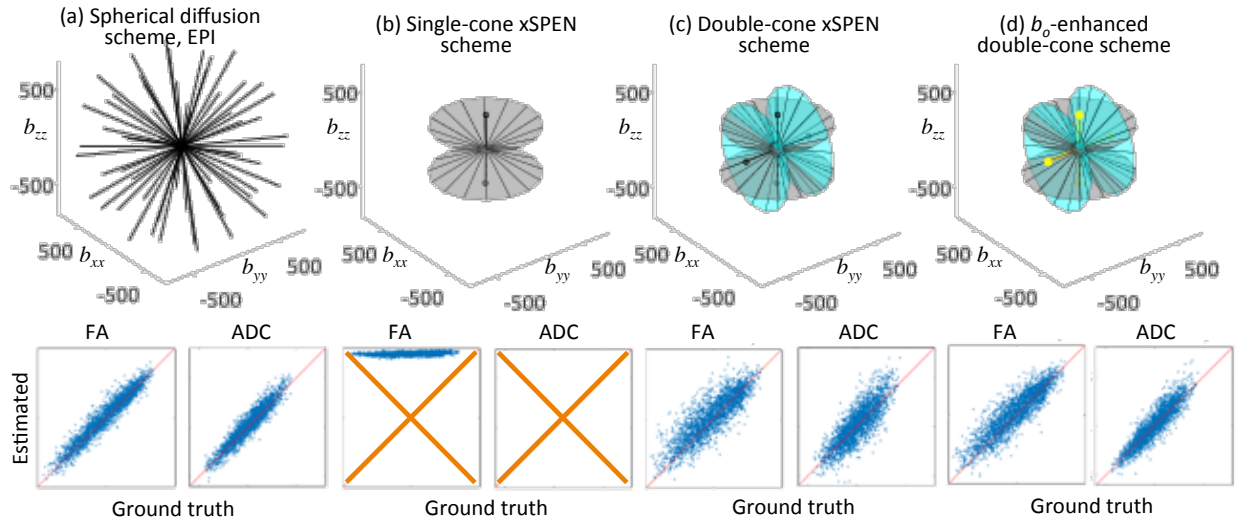

**Figure 2:** *b*-space sampling and reliability tests of various single-shot diffusion MRI gradient schemes. (a) Standard 30-directions spherical PGSE scheme applied on EPI (as supplied by the Siemens TrioTIM® scanner based on Ref. (39)), resulting in an excellent fit between the ground truth and the extracted values for an array of input fractional anisotropy (FA) and apparent diffusion coefficient (ADC) values ($r^2$=0.96, 0.95 respectively; thin red lines on these plots have a slope of 1). (b) Porting the same strategy (with 15 directions) to xSPEN leads to a single-cone *b*-space sampling, resulting in poor fits for either FA or ADCs ($r^2$=0.35, 0.15 respectively). (c) xSPEN's double-cone gradient scheme whereby bipolar diffusion gradients $G_d$ are applied along two sets of orthogonal orientations in independent experiments, providing a *b*-space sampling that reasonable estimates the FAs and ADCs ($r^2$=0.87, 0.85). (d) Enhanced double-cone gradient scheme incorporating 30 directions and independent $b_o$ estimations per cone, providing FA and ADC accuracies ($r^2$=0.89, 0.91) comparable to those achieved by EPI. See Methods for further details and for simulation parameters.

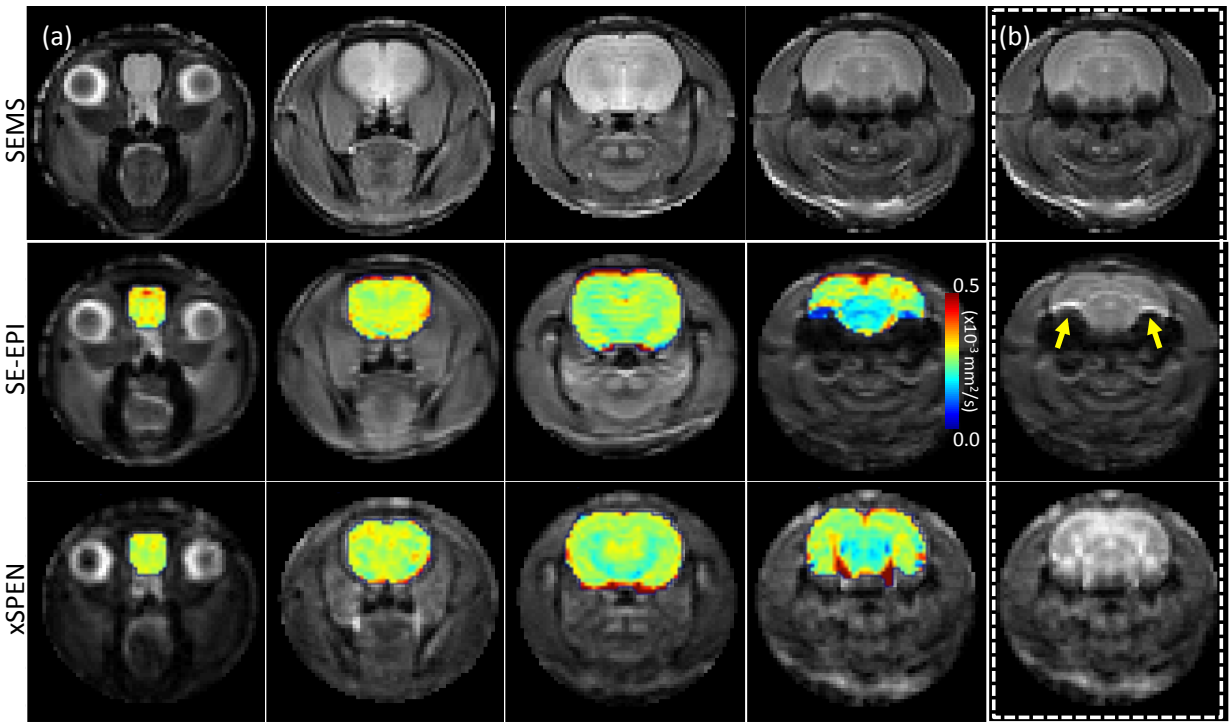

**Figure 3.** (a) Axial DWI maps (color) collected on an *ex-vivo* rat head. Top panel: SEMS magnitude images serving as anatomical reference for four representative slices. Middle panel: SE-EPI ADC maps overlaid on their corresponding $b_o$ magnitude images. Bottom panel: xSPEN ADC maps overlaid on their corresponding $b_o$ magnitude images (xSPEN acquisition running from top to bottom, explaining the decreased intensities towards bottom of the sample). (b) Comparison between various $b_o$ magnitude images acquired by the three methods; yellow arrows highlight brain inhomogeneity artifacts. Data were collected on a pre-clinical 7T scanner using six *b*-values with $|G_d| \leq 26.0$ G/cm; see Methods for additional scanning parameters.

Given these circumstances, we investigated alternative strategies for enabling reliable human diffusivity measurements using xSPEN. To this end an imaging scheme was devised, whereby the low-bandwidth xSPEN dimension was decoded twice, in two independent single-shot experiments that alternated the gradients employed along the RO and SS axes (Fig. 1b, purple labels). As a result of this the dominating *b*-weight associated with the SS axis also alternated among different (e.g., axial and sagittal) orientations. Figures 2c and 2d illustrate how this leads to two sets of orthogonal conic shapes in the sampling of the $\bar{\bar{b}}$-tensor space, rotated by 90° about $b_{yy}$. While not involving the comprehensive exploration of the tensorial space that is possible in SE-EPI, these two orthogonal xSPEN acquisitions lead to a sufficiently wide sampling to reliably measure the diffusion parameters. The lower panels in Figure 2 demonstrate this with a series of fractional anisotropy (FA) and apparent diffusion coefficient (ADC) plots employed to assess the reliability of the different $\bar{\bar{b}}$-space sampling schemes. These plots summarize a series of simulations where various 3D media (2601 "tissues") were assumed, each of them having a random proton density and realistic FA and ADC values (see Supporting Information and figure captions for further details). The signals arising from these synthetic "tissues" under the action of the different gradient schemes and pulse sequences were then calculated, and employed to estimate FA and ADC values on the basis of $S(\bar{\bar{b}})/S(0) = \exp(-\sum_{i,j} D_{ij} b_{ij})$, where $\{D_{ij}\}_{i,j=1-3}$ represent the diffusion tensor elements (40). $r^2$-values were then calculated against the ground truth FAs and ADCs known for the various "tissues", and from these the reliability of the various approaches was estimated. The most reliable ADC/FA assessments arose in all cases from SE-EPI (Fig. 2a); by contrast, a similar PGSE-based strategy gave unreliable results when incorporated into the original xSPEN sequence (Fig. 2b). Switching to two orthogonal acquisitions where the roles of the RO and SS axes in the xSPEN process are swapped and their outcomes processed in a combined analysis increased this considerably (Fig. 2c); the reliability of this double-cone $\bar{\bar{b}}$-space acquisition could be further enhanced by incorporating into the fits, independent $b_o$-samplings (i.e., samplings with all $G_d$s set to null) for each of the imaging schemes (Fig. 2d). The Extended Data Figure S2 illustrates an experimental validation of the resulting $b_o$-including "double-cone" approach, conducted on a water phantom in a clinical 3T MRI machine, showing essentially the same reliability as EPI-based maps. This latter gradient scheme was adopted for the human ADC and DTI xSPEN mapping.

*Diffusion in humans by xSPEN MRI.* Figure 4 compares SE-EPI and xSPEN results obtained on two representative human head slices, showing the $b_o$ magnitude images, ADC, FA and cFA (color-coded FA) maps to which each experiment leads. For an upper, relatively homogeneous brain slice (Fig. 4a), both methods show similar tissue contrasts and diffusivity information. Still, the FA maps show a slightly stronger contrast for SE-EPI vis-à-vis xSPEN –perhaps reflecting the different ADC mapping accuracies of the two methods as discussed in the context of Figure 2. For the lower brain slice (Fig. 4b), however, the SE-EPI data suffers from evident inhomogeneity distortions, clearly seen in the sinus regions and in the shapes of the eyes (yellow arrows). Interestingly, these are not only reflected in the SE-EPI $b_o$ images but also translate in diffusion map artifacts –for instance, in false anisotropies that SE-EPI describes in the center of the eyes's vitreous cFA maps.

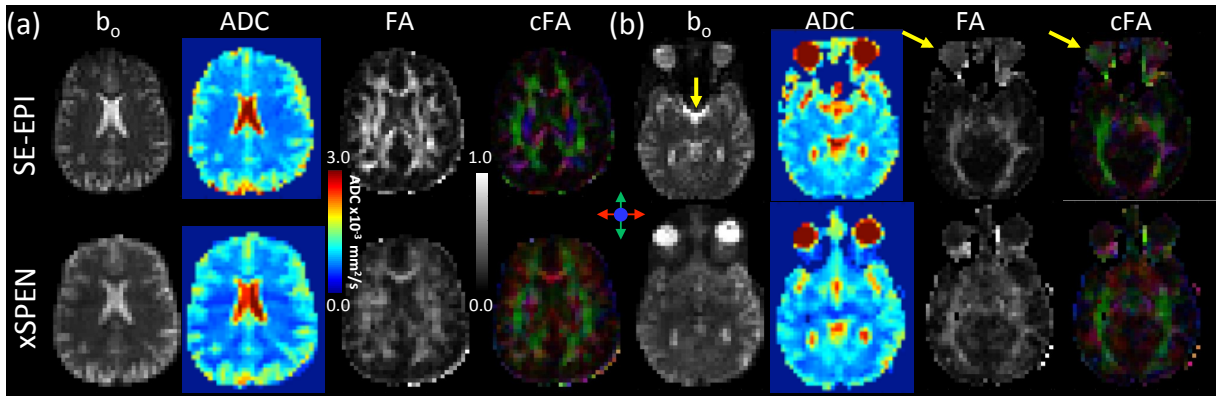

**Figure 4.** DTI datasets arising from SE-EPI and xSPEN for relatively homogeneous (a) and more challenging (b) head slices. For all cases these axial cuts show $b_o$, ADC, FA and colored-coded FA (cFA) with directions as defined by arrows in the figure's center. Yellow arrows highlight inhomogeneity artifacts, including a certain FA in the vitreous humor. xSPEN acquisitions run along an anterior-posterior axis, but do not show the intensity distortions remarked in Fig. 3 due to the longer T2s and weaker gradients involved in these human scans. Data were collected on a 3T human scanner using $|G_d|$=3.26G/cm; see Methods for additional scanning parameters.

Figure 5 compares another set of axial, coronal and sagittal human head images, showing SE-EPI and xSPEN ADC maps collected at 4mm isotropic resolution to probe diffusivity on head regions challenged by field inhomogeneity effects. Some of these regions are highlighted in the Figure by yellow arrows and include (a) the eyes, (b, c and e) the nose and nasal cavity, (d and g) the cerebellum, (f) the brain stem and (h) the tongue area. In SE-EPI these regions are partially or fully distorted and the ensuing ADC maps convey limited useful information, whereas xSPEN makes them clearly accessible. Worth comparing among these sets are the results observed for the vitreous humor, which in the center of the eyes include average ADCs of $3.7\pm0.2\times10^{-3}$ mm$^2$/s and $3.6\pm0.4\times10^{-3}$ mm$^2$/s for xSPEN and SE-EPI, respectively. These values are in good agreement with each other, as well as with literature reports (41).

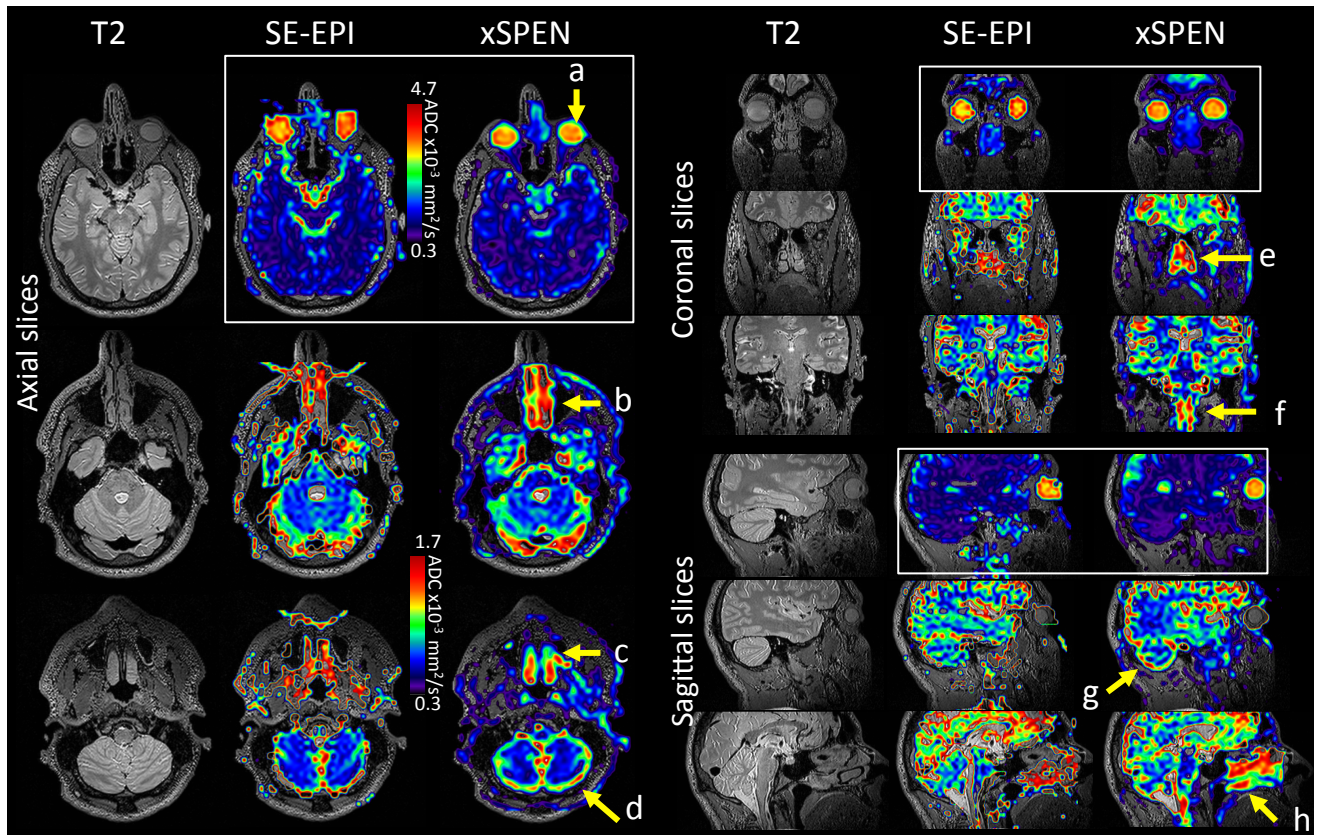

**Figure 5.** SE-EPI and xSPEN diffusion maps overlaid on T2 anatomical images. Yellow arrows highlight challenging head regions including the eyes (a), the nose and nasal cavity (b,c and e), the cerebellum (d and g), brain stem (f), and the tongue area (h). Notice that maps are given for two scales, with the larger ADC range (0.3-4.7x10$^{-3}$mm$^2$/s) applying to the framed white squares that include the eyes. In all cases the xSPEN axis run along the anterior-posterior direction; see Methods for additional scanning parameters.

Figure 6 shows an additional example collected at a higher, 3mm isotropic resolution, showing zoomed-in DTI maps of the head's anterior region. This folding-free zoom-in is possible at no cost in the sequence's complexity, and it helps to highlight features associated with the frontal lobes and eyes (Figure 6a). It also shows how xSPEN could help to target the diffusivity of the optical nerves, for which xSPEN FA and ADC maps were collected at a slight tilt (Fig. 6b) that allowed us to observe regions with distinct diffusion values in the center of these nerves. These features include a more mobile core with average ADC and FA values of $1.5\pm0.3\times10^{-3}$ mm$^2$/s and $0.25\pm0.09$ respectively, and a more slowly diffusing periphery with average ADC and FA values of $1.1\pm0.3\times10^{-3}$ mm$^2$/s and $0.28\pm0.10$ respectively. Judging by the corresponding T1-weighted images, these data are affected by minimal artifacts despite the strong air-interface $\Delta B_o$s associated to these tissue regions. These ADC values are in agreement with literature reports based on fast spin-echo and on EPI measurements (0.8-1.4x10$^{-3}$mm$^2$/s), even if reported FA values for optic nerves have been usually higher (0.39-0.64) (42,43). The origin of this variation is under investigation.

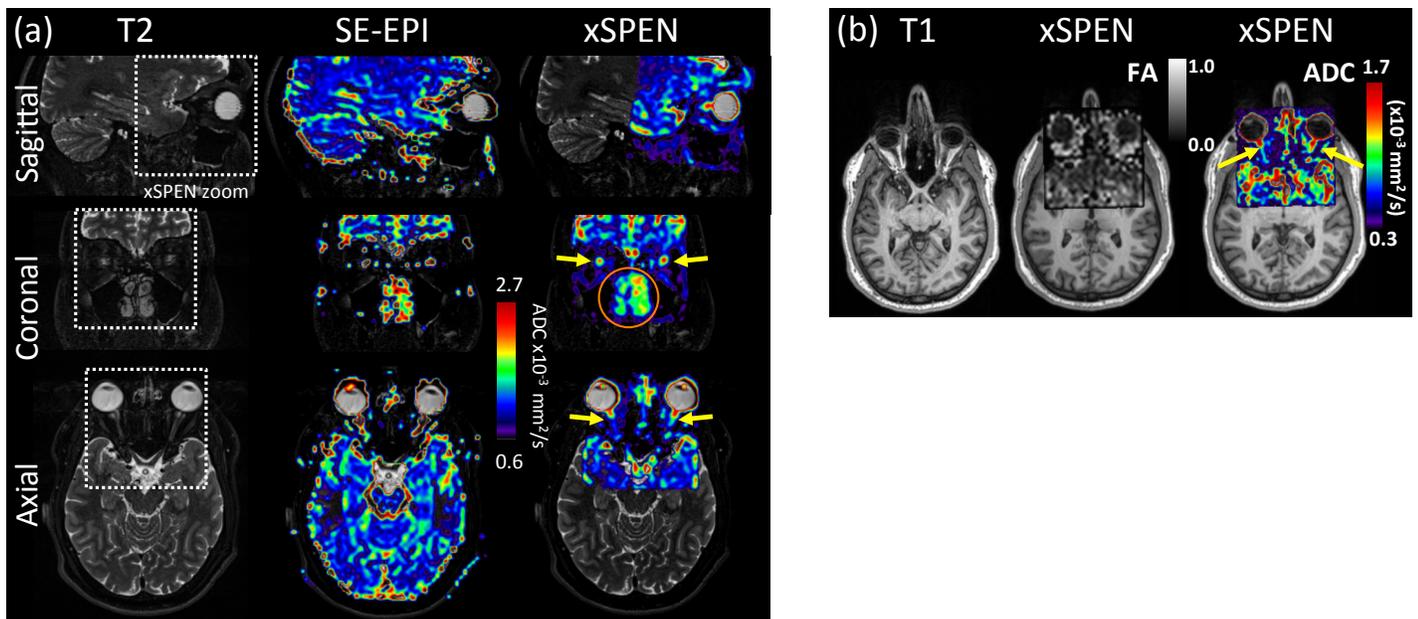

**Figure 6.** Zoomed-in xSPEN diffusion data arising from the frontal lobe (dashed square regions) containing the optic nerve indicated by the yellow arrows. (a) SE-EPI and xSPEN ADC maps overlaid on T2-weighted anatomical images, showing the optic nerve in three different orientations. Notice the regions missing in the EPI maps owing to susceptibility distortions, including the nasal cavity (orange circle). (b) FA and ADC maps obtained with xSPEN zooms that were slightly tilted to better align along the optic nerve, overlaid on anatomical T$_1$-weighted images. The ADC scale has been adapted to highlight the nerves. In all cases the xSPEN axis run along the anterior-posterior direction; see Methods for additional scanning parameters.

## Discussion and Conclusions

Different diffusion-monitoring variants have been explored over the years as diffusion-monitoring alternatives to EPI, particularly within the context of overcoming heterogeneous field environments. These include multi-shot schemes liable to motion-derived errors, as well as single-shot acquisitions based on radial samplings or on the use of multiple rf-driven spin echoes (15-19). Recently introduced single-shot 2D SPEN methods can achieve many of these goals (21,35,44-46), yet if overwhelmed by inhomogeneities even these methods are liable to yield corrupted images reflecting $\Delta B_o$ displacements. xSPEN by contrast shows unprecedented resilience to in-plane field inhomogeneities, with the sole distortions identified in their single-scan 2D images arising from limitations in the slice selection process (30). xSPEN's immunity to field inhomogeneities stems from its reliance on a constantly active frequency broadening mechanism that incorporates background inhomogeneities into the image-formation process. This use of a continuous gradient throughout the encoding and acquisition, however, results in a strong intrinsic *b*-weighting that needs to be accounted for when attempting quantitative diffusion measurements. In order to enable such accounting, the sequence's diffusion effects were quantitatively evaluated using a customized, spatially-localized *b*-matrix analysis. With the aid of this formalism and of numerical simulations on "synthetic tissues", a diffusion-weighting scheme was devised that overcomes xSPEN's original limitations; this operates by exchanging, in independent measurements, the roles that orthogonal gradients play as the broadening mechanism enabling the xSPEN image formation. Phantom experiments validated the quantitativeness of the resulting "double-cone" *b*-sampling approach; when the diffusion gradients were not sufficiently intense, a double PGSE block could be readily incorporated into the sequence. When implemented on human volunteers (Figs. 4-6) the ensuing double-PGSE, double-cone scheme improved the ADC and FA depiction afforded by SE-EPI, providing reliable maps of isotropic and anisotropic diffusivity for the brain stem, cerebellum, mouth and the ocular regions. Thanks to its reliance on frequency swept pulses xSPEN also provided a built-in "zooming" capability that allowed us to focus on the frontal brain region, where ADC mapping of the optical nerve was demonstrated in all three orientations –always evidencing a good, rounded shape and faithful mapping of the nerves' diffusivity characteristics to their original locations.

Despite these features, xSPEN still exhibits a number of limitations that remain to be overcome. Foremost among these is its SNR limitation; while we have managed to overcome these for the original SPEN experiment using super-resolution procedures (47,48), similar procedures remain to be devised for xSPEN. In their absence, we could not explore higher human spatial resolutions than 3mm isotropic. An additional limitation of xSPEN compared to other single-shot methods –included its SPEN predecessors– is its higher SAR. This results from xSPEN's use of two frequency-swept inversion pulses, which led to us reach ca. 90% of the maximum allowed (normal mode) SAR values for performing the multi-slice experiments illustrated above. Approaches to minimize this problem's impact are also under investigation. Despite these two main limitations we believe that xSPEN's ability to target challenging inhomogeneity-dominated regions, of the kind that are usually inaccessible by traditional single-shot methods, can open new opportunities in basic and in clinical investigations. This immunity to field heterogeneities was here demonstrated for various head regions of healthy volunteers; we are exploring what new avenues can be opened when imaging other body regions, as well as tissues proximate to metal implants including spine, mouth and dental MRI.

**Acknowledgments.** We are grateful to Dr. Sagit Shushan (Wolfson Medical Center) and the Weizmann MRI team (Edna Furman-Haran, Fanny Attar and Nachum Stern) for assistance in the human scans, and to Dr. Qingjia Bao (Weizmann) for assistance in the preclinical experiments. ZZ thanks Israel's Council of Higher Education and to the Koshland Foundation for partial postdoctoral fellowships. Financial support from the Israel Science Foundation grant 795/13, the EU through ERC-2016-PoC grant #751106, Minerva funding (#712277) from the Federal German Ministry for Education and Research, the Kimmel Institute for Magnetic Resonance and the generosity of the Perlman Family Foundation, are also acknowledged.

**Author Contributions:** ES, GL, ZZ & LF designed research; ES, GL, ZZ acquired the data; ES, GL contributed new tools and analyzed data; ES & LF wrote the paper.

**Conflict of interests:** The authors declare no competing financial interests.

## Methods

*Preclinical scans.* Phantom and *ex-vivo* whole-head rat experiments were performed on a DD2® 7T/110mm horizontal magnet scanner (Agilent Technologies, Santa Clara, CA) using a quadrature-coil probe. For these scans the xSPEN pulse sequence shown in Figure 1a was used and compared against a SE-EPI sequence with similarly structured timing and gradient strengths. These sequences ran in and were processed within Agilent's VNMRJ 3.2 software environment. For anatomical reference, scanner-provided fast spin-echo multi-shot (SEMS) sequences were used. Scanning parameters for the Titanium-Lego® phantom included TE≈50ms, $T_a$=22.0ms, FOV=32x32mm², resolution=0.5×0.5mm², 3.0mm slice. Number of averages: SEMS=1, SE-EPI=1 SPEN=1 and xSPEN=2. TR: SEMS=2s, SE-EPI=8s, SPEN=8s, xSPEN=8s. Additional parameters: SPEN's encoding bandwidth BW=18.0kHz, $G_y$=1.33G/cm; xSPEN's encoding bandwidth BW=5.8kHz, $G_y$=0.21G/cm, $G_z$=2.27G/cm. The DWI parameters for all phantom experiments were $\delta$=3ms, $\Delta$=12ms, maximum diffusion gradients $G_d$=35G/cm applied in three orthogonal directions leading to a $b_{max}$=870s/mm$^2$, six $b_{max}$-scaling values (0, 0.25, 0.63, 0.77, 0.89 and 1.0). For the *ex vivo* DWI experiments, xSPEN and SE-EPI parameters were $\delta$=5ms, $\Delta$=13ms, maximum diffusion gradients $G_d$=26G/cm applied in three orthogonal directions leading to a $b_{max}$=1370s/mm$^2$, six $b_{max}$-scaling values (0, 0.25, 0.63, 0.77, 0.89 and 1.0). Common scanning parameters included TR=5s, FOV=32x32 mm$^2$, resolution=0.5×0.5mm$^2$, 3.0mm slice thickness. Number of averages: SEMS=2, SE-EPI=4, xSPEN=4. Additional parameters: for SE-EPI TE=40ms and $T_a$=22. ms; for xSPEN middle TE=62ms, $T_p$=$T_d$/2=11.0ms, encoding bandwidth BW=3.6kHz, $G_y$=0.13G/cm, $G_z$=1.42G/cm.

*Clinical scans.* Phantom and *in vivo* human head diffusion maps were acquired at 3T using a Siemens TrioTIM® scanner (Erlangen, Germany) equipped with a 32-channels head coil. Three subjects (males aged 26, 28 and 30) were scanned for improving the optic nerve and ocular globe diffusional statistics; all experiments were approved by the Internal Review Board WOMC-0091-11 of the Wolfson Medical Center (Holon, Israel) and collected after obtaining informed suitable consents.. The phantom consisted of a plastic bottle containing 1.9L of water with 3.75g $NiSO_4*6H_2O$ + 5g NaCl. For these scans the diffusion xSPEN pulse sequence in Figure 1b was used, and compared against a double bipolar diffusion SE-EPI sequence (39). For the simulations (Figs. 2 and S1), the water phantom experiments (Fig. S2) and the lower-resolution human head scans (Figs. 4 and 5), the $b_{max}$ used for both xSPEN and SE-EPI had a nominal 1000 s/mm$^2$ value. Additional xSPEN diffusion parameters included $\delta$=16ms, $\Delta$=57ms, $G_d$=3.26G/cm. xSPEN scanning parameters for these data were: 48 slices, one nominal |b|=$b_{max}$ value, $T_p$=$T_d$/2=10.9ms, $G_y$=0.027G/cm, $G_z$=1.24 G/cm, chirp bandwidth $BW$=2.2kHz, FOV=185x193mm$^2$ (PExRO), resolution=4.0mm$^3$ isotropic, TR=25s, SAR=90%. Number of averages: SE-EPI=1, xSPEN=2. Other SE-EPI scanning parameters: TE=80ms, TR=10s, FOV=184x256mm$^2$ (PExRO), resolution=4.0mm$^3$ isotropic. For the higher, 3.0mm$^3$ isotropic resolution head scans (Fig. 6) the $b_{max}$-value for both xSPEN and SE-EPI was 800s/mm$^2$. Additional xSPEN parameters: $\delta$=17.2ms, $\Delta$=49.1ms, $G_d$=2.95G/cm, 32 slices, one b-value, TR=15s, FOV=90x96mm$^2$ (PExRO), $T_p$=$T_d$/2=7.75ms, $G_y$=0.0504G/cm, $G_z$=1.52 G/cm, chirp bandwidth $BW$=2kHz, SAR=90%.

Additional scanning parameters: Number of averages: SE-EPI=1, xSPEN=4. Other SE-EPI scanning parameters: TE=84ms, TR=10s, FOV=180x252mm$^2$ (PExRO), resolution=3.0 mm$^3$ isotropic. T1- and T2-weighted scans are also included as anatomical references.

# Figure Captions

**Figure 1.** (a,b) xSPEN diffusion pulse sequences assessed in this study, involving in all cases a slice-selective $90^0$ excitation pulse ($p_1$) followed by two $180^0$ chirped pulses. In (a), a single diffusion-weighting PGSE block is placed in the pre-encoding $(T_a+p_1)/2$ delay required for targeting the desired FOV under full-refocusing. In (b) two PGSE diffusion blocks are placed on both sides of the $180^0$ chirped pulses to enable a larger diffusion weighting, and both RO and SS axes are alternated among *x* and *z* orientations in order to overcome the otherwise dominating $b_{zz}$-weighting derived from the $G_z$ gradient. The RF/ADC line displays the pulses and signal acquisition; RO, SPEN and SS display orthogonal gradient directions; $G_d$ are diffusion-weighting gradients of duration δ and stepped amplitudes (in grey); $G_{pr}$, purge gradients; $G_{ro}$, readout acquisition gradients; $T_a$, acquisition time. (c) MRI results obtained on a preclinical 7T scanner for an agar phantom containing a titanium screw and a Lego® piece arranged as indicated on the left, imaged by a reference multi-shot spin echo and by several (SE-EPI, SPEN, xSPEN) single shot methods. In the latter case, the sequence in panel (a) was used with $G_d$=0. Notice that while metal-induced field distortions are evident in all four methods, the xSPEN results most faithful reproduce the original phantom distribution and provide the most reliable ADC map.

**Figure 2.** *b*-space sampling and reliability tests of various single-shot diffusion MRI gradient schemes. (a) Standard 30-directions spherical PGSE scheme applied on EPI (as supplied by the Siemens TrioTIM® scanner based on Ref. (39)), resulting in an excellent fit between the ground truth and the extracted values for an array of input fractional anisotropy (FA) and apparent diffusion coefficient (ADC) values ($r^2$=0.96, 0.95 respectively; thin red lines on these plots have a slope of 1). (b) Porting the same strategy (with 15 directions) to xSPEN leads to a single-cone *b*-space sampling, resulting in poor fits for either FA or ADCs ($r^2$=0.35, 0.15 respectively). (c) xSPEN's double-cone gradient scheme whereby bipolar diffusion gradients $G_d$ are applied along two sets of orthogonal orientations in independent experiments, providing a *b*-space sampling that reasonable estimates the FAs and ADCs ($r^2$=0.87, 0.85). (d) Enhanced double-cone gradient scheme incorporating 30 directions and independent $b_o$ estimations per cone, providing FA and ADC accuracies ($r^2$=0.89, 0.91) comparable to those achieved by EPI. See Theory and Methods for further details and for simulation parameters.

**Figure 3.** (a) Axial DWI maps (color) collected on an *ex-vivo* rat head. Top panel: SEMS magnitude images serving as anatomical reference for four representative slices. Middle panel: SE-EPI ADC maps overlaid on their corresponding $b_o$ magnitude images. Bottom panel: xSPEN ADC maps overlaid on their corresponding $b_o$ magnitude images (xSPEN acquisition running from top to bottom, explaining the decreased intensities towards bottom of the sample). (b) Comparison between various $b_o$ magnitude images acquired by the three methods; yellow arrows highlight brain inhomogeneity artifacts. Data were collected on a pre-clinical 7T scanner using six *b*-values with $|G_d| \leq 26.0$ G/cm; see Methods for additional scanning parameters.

**Figure 4.** DTI datasets arising from SE-EPI and xSPEN for relatively homogeneous (a) and more challenging (b) head slices. For all cases these axial cuts show $b_o$, ADC, FA and colored-coded FA (cFA) with directions as defined by arrows in the figure's center. Yellow arrows highlight inhomogeneity artifacts, including a certain FA in the vitreous humor. xSPEN acquisitions run along an anterior-posterior axis, but do not show the intensity distortions remarked in Fig. 3 due to the longer T2s and weaker gradients involved in these human scans. Data were collected on a 3T human scanner using $|G_d|$=3.26G/cm; see Methods for additional scanning parameters.

**Figure 5.** SE-EPI and xSPEN diffusion maps overlaid on T2 anatomical images. Yellow arrows highlight challenging head regions including the eyes (a), the nose and nasal cavity (b,c and e), the cerebellum (d and g), brain stem (f), and the tongue area (h). Notice that maps are given for two scales, with the larger ADC range (0.3-4.7x10$^{-3}$mm$^2$/s) applying to the framed white squares that include the eyes. In all cases the xSPEN axis run along the anterior-posterior direction; see Methods for additional scanning parameters.

**Figure 6.** Zoomed-in xSPEN diffusion data arising from the frontal lobe (dashed square regions) containing the optic nerve indicated by the yellow arrows. (a) SE-EPI and xSPEN ADC maps overlaid on T2-weighted anatomical images, showing the optic nerve in three different orientations. Notice the regions missing in the EPI maps owing to susceptibility distortions, including the nasal cavity (orange circle). (b) FA and ADC maps obtained with xSPEN zooms that were slightly tilted to better align along the optic nerve, overlaid on anatomical T$_1$-weighted images. The ADC scale has been adapted to highlight the nerves. In all cases the xSPEN axis run along the anterior-posterior direction; see Methods for additional scanning parameters.

Extended Data for

# Diffusion MRI measurements in challenging head and brain regions via cross-term spatiotemporally encoding

Eddy Solomon[a], Gilad Liberman[a], Zhiyong Zhang and Lucio Frydman*

*Department of Chemical Physics, Weizmann Institute of Science, Rehovot 76100, Israel*

## A. Theoretical analysis of diffusion in xSPEN MRI.

To provide a quantitative framework capable of estimating the effects that diffusion will have in xSPEN MRI, we rely on a formalism whereby the local derivatives of the spin evolution phases are first calculated, and then combined with the diffusion model discussed by Karlicek and Lowe (1) according to which the diffusion-derived attenuation imposed by gradients throughout an NMR sequence is summarized by a *b*-value

$$b(t) = \gamma^2 \cdot \int_0^t K^2(t')dt' \qquad [S1]$$

where $K(t') = \int_0^{t'} G(t'')dt''$ is a wavenumber encompassing the action of all gradients up to a particular time *t'*. As shown in (2,3), the application of frequency swept pulses under the action of gradients as done in SPEN/xSPEN, results in a spin dephasing which, by contrast to the assumptions leading to Eq. [S1], is neither linear in space, nor independent of position. To account for this we preserve Eq. [S1] but re-express the *K*-wavenumber in terms of a local spatial dispersion $K^{local}$, describing the dephasing experienced by the spins within a neighboring region that is relevant in terms of the diffusion length scale (2,4). A Taylor expansion allows one to describe this wavenumber in proximity to an arbitrary $r_0$ as $K^{local}(t', r_o) = \dfrac{d\varphi(t', r_o)}{dr}$, an expression that becomes identical to the Karlicek-Lowe formulation if $\varphi$'s dephasing has been imparted solely by a linear gradient.

Using this formalism for calculating the diffusion-driven signal attenuation as a function of time and position, the decays expected for the xSPEN sequences described in Figure 1 of the main text, were estimated. For simplicity, we only took into consideration the imaging gradients along the xSPEN-relevant (y,z)-axes, disregarding the effects of the rapidly-oscillating RO (x-axis) gradient, and assuming that the diffusion gradients (greyed $G_d$s in Fig. 1) were initially null. The relevant manipulations therefore include a slice-selective 90° excitation followed by two identical frequency-swept inversion pulses acting in synchrony with a bipolar $\pm G_y$ –all of this imparted while in the presence a constant $G_z$, which stays active throughout the course of an acquisition lasting a duration $T_a$. Referring to $t_-$ as the time in which spins positioned at a given $r_o=(y_o,z_o)$ coordinate are addressed by the first inversion pulse, $t_+$ as the time when these spins are addressed by the second pulse, and $T_e=T_a/2$ as the duration of these pulses, the formalism described in Refs (2,5) allows one to compute the local wavenumbers accumulated by the spins during the course of the xSPEN encoding and up to the instant of their subsequent acquisition as

$$K^{local}(t',r_o) = \begin{cases} -G_y t + G_z t, & 0 \le t \le t_- \\ G_y t - G_z t, & t_- \le t \le T_e \\ G_y t - G_z t, & T_e \le t \le T_e + \dfrac{T_a}{2} \\ \dfrac{G_y}{2}(-2t + T_a + 4T_e) - G_z t, & T_e + \dfrac{T_a}{2} \le t \le T_e + \dfrac{T_a}{2} + t_+ \\ G_y t - \dfrac{G_y T_a}{2} + G_z(t - T_a - 2T_e), & T_e + \dfrac{T_a}{2} + t_+ \le t \le T_e + \dfrac{T_a}{2} + T_e \\ 2G_y T_e + G_z t - \dfrac{G_z T_a}{2}, & 0 \le t \le T_a \end{cases} \qquad [S2]$$

Using such dynamic evolution wavenumbers, the *b*-values can be calculated from Eq. [S1] in a similar piece-wise temporal fashion as

$$b(t) = \gamma^2 \begin{cases} \dfrac{(G_y^2 + G_z^2)t^3}{3}, & 0 \le t \le t_- \\ -\dfrac{(G_y^2 + G_z^2)(t_-^3 - T_e^3)}{3}, & t_- \le t \le T_e \\ \dfrac{T_a(12G_y^2 T_e^2 + G_z^2(T_a^2 + 6T_a T_e + 12T_e^2))}{24}, & T_e \le t \le T_e + \dfrac{T_a}{2} \\ \dfrac{t_+[4G_y^2(t_+^2 - 3t_+ T_e + 3T_e^2) + G_z^2(3T_a^2 + 6T_a(t_+ + 2T_e) + 4(t_+^2 + 3t_+ T_e + 3T_e^2))]}{12}, & T_e + \dfrac{T_a}{2} \le t \le T_e + \dfrac{T_a}{2} + t_+ \\ -\dfrac{G_y^2 t_+^3}{3} - G_y^2 t_+^2 T_e - G_y^2 t_+ T_e^2 + \dfrac{7G_y^2 T_e^3}{3} + \dfrac{G_z^2(-T_a^3 + (T_a - 2t_+ + 2T_e)^3)}{24}, & T_e + \dfrac{T_a}{2} + t_+ \le t \le T_e + \dfrac{T_a}{2} + T_e \\ \dfrac{G_z^2 T_a^3}{12} + 4G_y^2 T_a T_e^2 & 0 \le t \le T_a \end{cases} \qquad [S3]$$

From this equation the diffusion-driven decay $S(b)/S(0) = \exp[-b \cdot D]$ affecting xSPEN even in the absence of diffusion-sensitizing gradients –where $D$ is an isotropic diffusion coefficient and $S(0)$ is the signal in the absence of diffusion– follows.

In general, displacement measurements require extending the above calculation to account for the presence of diffusion gradients $G_d$ applied along multiple, non-coincident spatial orientations (6-8). These gradients can be introduced in the xSPEN scheme as illustrated in the manuscript's Fig. 1a, which places a PGSE block during a built-in $(T_a+p_1)/2$ free evolution delay introduced for the sake of achieving full refocusing. These pulsed magnetic field gradients $G_d$, acting along arbitrary orientations for a duration δ and separated by a diffusion-sensitizing time Δ, can be accounted by extending the $K^{local}$-based formalism that lead to Eq. [S2], to the tensorial form (9)

$$\bar{\bar{b}}(t,r_0) = \gamma^2 \int_0^t (\vec{\nabla}(\varphi(t',r_0))\vec{\nabla}(\varphi',r_0)^T)dt', \qquad [S4]$$

where $\bar{\bar{b}}$ is now a matrix dictated by products of the localized phase derivatives of $\varphi(t',r_0)$, given by both the imaging and the diffusion gradients. The complexities of this behavior are further illustrated in Extended Data Figure S1, which describes how the various tensorial b-components vary for different positions along the imaged axis over the course of an xSPEN acquisition.

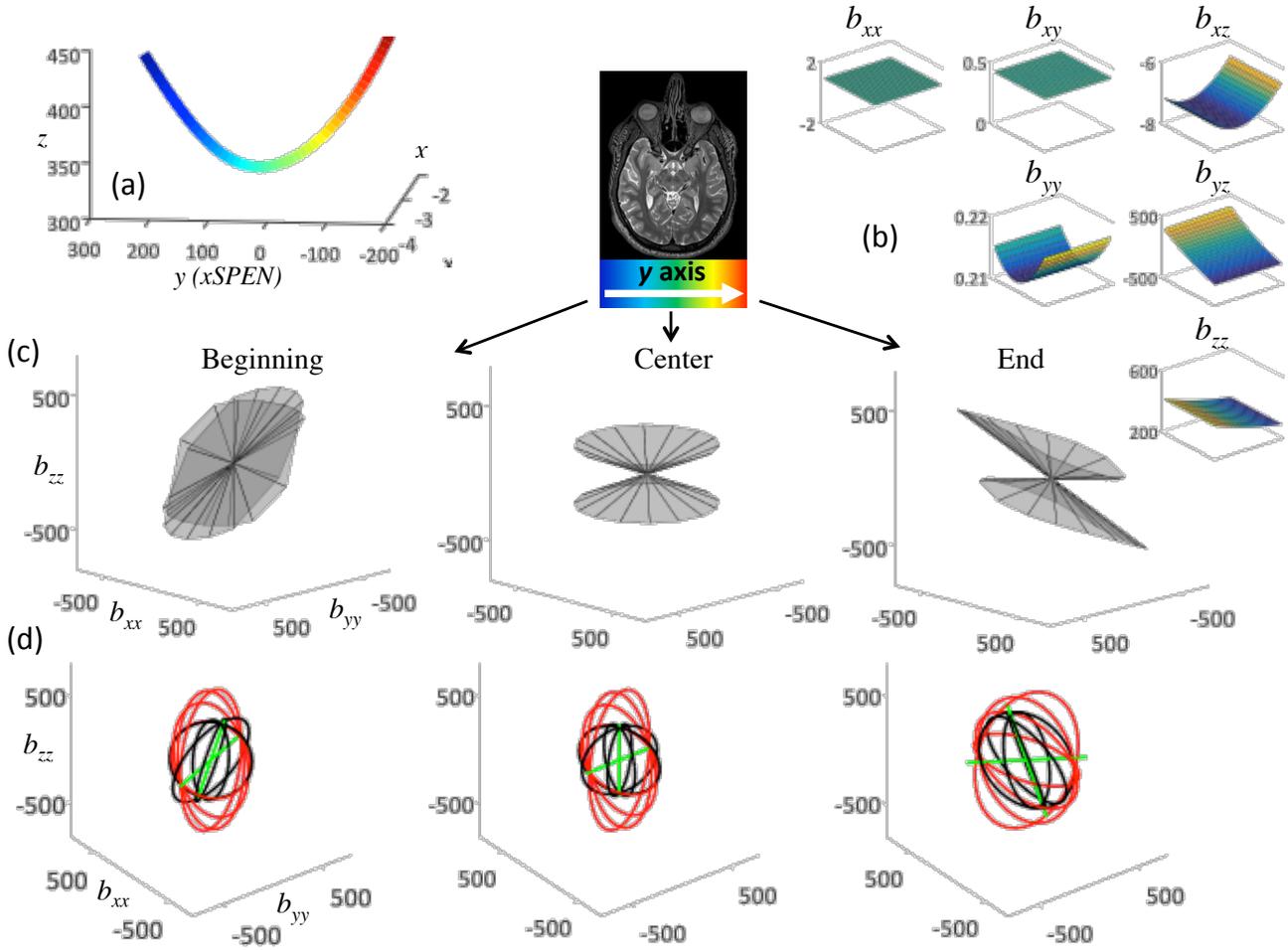

**Extended Data - Figure S1.** Features associated to xSPEN's use in diffusion measurements, illustrated with results computed on the basis of Eqs. [S1]-[S4]. For all axes in all panels, values are given in units of s/mm$^2$. (a) Curve connecting the tips of the major $\bar{\bar{b}}$-matrix eigenvectors arising in the absence of diffusion-sensitizing gradients, as a function of positions along the y (xSPEN) axis; colors correspond to progress in the decoding along the white arrow shown under the brain image. In correspondence with the acquisition parameters used for the experiments in the main text's Fig. 4, the spatial dependency of these b-values is strong along the y-axis, sizable but nearly constant along the z-axis (~400 s/mm$^2$), and negligible along the x-axis. (b) Alternative rendering of the same information, showing a strong weighting of all $\bar{\bar{b}}$-matrix components associated with the x/y/z axes. (c) Description of how the largest b-tensor elements vary over the course of a xSPEN acquisition involving PGSE gradients rotating solely in the (x,y)-plane, showing the position-dependent distortions introduced by the slice-selection gradient. (d) Coverage of the $\bar{\bar{b}}$-matrix components afforded by the double-cone gradient scheme (Fig 2c), focusing on the orientations of the major eigenvectors arising as a function of y-position when choosing the SS axis along z (black ellipses) or along x (red ellipses). The green axes represent the two $b_o$ samples associated with these orthogonal choices.

### B. Gradient scheme validations.

The reliability of the various gradient schemes discussed in the main text's Figure 2 were tested on a set of 2601 "synthetic tissues", where each of these samples was assigned a random proton density and axially-symmetric diffusion tensors with randomized directionality and eigenvalues spanning realistic FA (0-1 arbitrary) and ADC (0.4-1.8x10$^{-3}$) values. The signals arising from these different tissues under the action of the various gradient schemes were simulated using the Bloch-Torrey formalism (10), and Gaussian noise was added to these calculated signals so that the mean $b_0$–image SNR would be 7%. Based on these synthetic sets, FA and ADC values were then estimated, and both mean absolute differences (which can be appreciated in Fig. 2 as deviations from the red-lined unity slopes graphed) and $r^2$ values, were calculated against the ground truth FAs and ADCs for the various tested gradient schemes. The Extended Data Figure S2 illustrates an experimental validation of the resulting $b_o$-including "double-cone" approach, conducted on a water phantom in a clinical 3T MRI machine, showing essentially the same reliability as EPI-based maps. This latter gradient scheme was adopted for the human ADC and DTI xSPEN mapping.

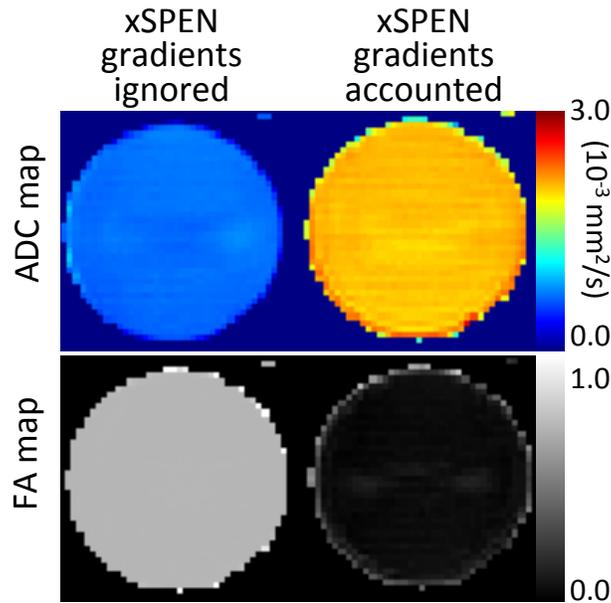

**Extended Data Figure S2.** Experimental validation of the "$b_o$-enhanced double-cone" gradient scheme introduced in the main text Figs. 1b and 2d, on a water phantom examined in a 3T clinical scanner. The xSPEN pulse sequence in Fig. 1b was used to derive the ADC and FA maps. In the left-hand column these were derived under the assumption that the diffusion-driven signal attenuation solely arise due to the effects of the $G_d$ diffusion gradients (11); in the right-hand column maps accounted for both the xSPEN imaging and the PGSE bipolar gradients as per the analytical calculation deriving from Eqs. [S1]-[S4].